Classification: PHYSICAL SCIENCES, Physics

# Estimation of critical exponents from the cluster coefficients: Application to hard spheres

E. Eisenberg[1] and A. Baram[2]

1. School of Physics and Astronomy, Raymond and Beverly Sackler Faculty of Exact Sciences, Tel Aviv University, Tel Aviv 69978, Israel

2. Soreq NRC, Yavne 81800, Israel

Corresponding author:

    Eli Eisenberg,
    School of Physics and Astronomy,
    Raymond and Beverly Sackler Faculty of Exact Sciences,
    Tel Aviv University,
    Tel Aviv 69978, Israel
    Tel: +972-3-640-7723
    Fax: +972-3-642-2979
    Email: elieis@post.tau.ac.il

16 pages, 2 figures, 2 tables

**Abstract**: For a large class of repulsive interaction models, the Mayer cluster integrals can be transformed into a tridiagonal real symmetric matrix $R_{mn}$, whose elements converge to two constants with $1/n^2$ correction. We find exact expressions, in terms of these correction terms, for the two critical exponents describing the density near the two singular termination points of the fluid phase. We apply the method to the hard-spheres model and find that the metastable fluid phase terminates at $\rho_t=0.751(5)$. The density near the transition is given by $\rho_t-\rho \sim (z_t-z)^{\sigma'}$, where the critical exponent is predicted to be $\sigma'=0.0877(25)$. Interestingly, the termination density is close to the observed glass transition, and thus the above critical behavior is expected to be associated with the onset of glassy behavior in hard spheres.

**PACS**: 64.70.Pf, 64.60.Fr, 61.20.-p,

Hard-core models have long played a central role as models for structural ordering transitions. These models are purely entropy-driven, thus capturing the essential molecular mechanism that drives freezing transitions. In particular, the equation of state of classical hard-spheres and their freezing transition have been extensively studied for over a century, dating back to the pioneering works by the founding fathers of statistical mechanics [1,2]. Nevertheless, there are still a lot of fundamental unresolved problems regarding this system. Early Monte-Carlo (MC) works have established that in three dimensions the system freezes through a first-order phase transition at a density $\rho_f=0.665(2)$ [3-6] (all densities here are scaled by the closest packing density; for hard spheres the volume fraction density $\hat{\rho}$ is given by $\hat{\rho} = \pi\rho/3\sqrt{2}$). Increasing the density of the hard-sphere fluid quickly enough, crystallization can be avoided, and the system stays in a metastable super-cooled fluid phase. As the density of super-cooled fluid increases, its dynamics becomes slower and slower. Experimental [7,8] and computational [9-11] studies have shown that the typical relaxation times increases very fast around $\rho_g \sim 0.76$. Despite its long history, the question whether this increase is due to a true thermodynamic glass transition or a purely dynamical phenomenon is still hotly debated [12-18].

The most pronounced manifestation (and, experimentally, the definition) of glassiness is through the dynamical properties. Thus, characterization of the onset of glassiness using statistical-mechanics approaches is challenging. Several approaches have been taken to meet this challenge, including the Mode-Coupling theory [19], using the replica method [16] and analysis of the virial expansion [20]. In this work we suggest to use the Mayer cluster expansion for describing the metastable super-cooled fluid and, in particular, the critical behavior near its termination point.

It was previously shown that the Mayer cluster expansion for the fluid density can be represented by a real symmetric tri-diagonal matrix R (see below, equations (5) and (6)) [21]. The matrix elements of R converge to constant terms A (off-diagonal) and B (diagonal), and this fact was previously used to obtain very good approximations for the fluid density [22,23]. However, this approximation fails near the fluid termination point, where it predicts a universal square-root critical behavior. In this work, we provide evidence that not only the matrix elements converge to a constant, but the asymptotic convergence of the matrix elements towards their asymptotic values is also universal – for a large set of hard-core models the deviation of the matrix elements from their limiting value scales like $1/n^2$. We analyze the effect of these correction terms on the critical behavior and obtain an exact expression for the critical exponents characterizing the singular z-dependence of the density near the singular points in terms of the correction terms which are readily extracted from the first few cluster integrals. The results are then applied to the hard-spheres model.

The Mayer cluster expansion provides a convenient low-density description in terms of the activity z. For systems with purely repulsive interactions, the cluster coefficients take a universal asymptotic form [24-26]. The thermodynamic potentials have two branch point singularities on the real z axis: one non-physical singular point on the negative axis at $-z_0$, and a second singularity at $z=z_t>z_0$ characterizing the termination of the (stable or metastable) fluid phase. The non-physical singularity at negative activity is known to be universal, i.e., the non-analytic z-dependence of the density near the singularity is $\rho_{sing}(z) \sim (z+z_0)^\sigma$, where the exponent $\sigma=\phi(d)-1$ is model-independent, and depends only on the spatial dimension [24-26].

It is always possible to transform the Mayer cluster series of the thermodynamic potentials into a symmetric tri-diagonal matrix R which provides the thermodynamic potentials. Moreover, for models with repulsive interaction only, the R matrix is a real symmetric one, with spectrum of real eigenvalues [21]. The reciprocals of the eigenvalues of this matrix are the Yang-Lee zeroes of the grand-canonical partition function, and lie on two intervals along the real activity axis: $z<-z_0$ and $z>z_t$. The critical properties are then determined by the two singular points closest to the origin on the negative ($-z_0$) and positive ($z_t$) real axis. The fact that all the eigenvalues of R are real simplifies considerably its treatment, and underlies the success of the present approach. Furthermore, the matrix elements of R asymptotically converge towards constant values B (diagonal) and A (near diagonal), which determine the location of the two branch point singularities of the fluid thermodynamic functions [21]. This behavior has been verified for all (>25) repulsive models available, without even a single exception. These features make the matrix representation very convenient as a starting point for analytical and numerical treatments.

In this work we provide evidence that the asymptotic convergence of the matrix elements towards their asymptotic values is also universal – for a large set of models the deviation of the matrix elements from their limiting value scales like $1/n^2$:

$$R_{nn} = B + b/n^2 \qquad R_{n,n+1} = A + a/n^2. \qquad (1)$$

We then find an exact relation between the coefficients of these asymptotic corrections and the critical exponents characterizing the singular z-dependence of the density near the two singular points – the non-physical, universal, singularity at negative activity, and the physical, model-dependent, singularity at positive activity. One thus obtains a simple recipe for calculating the critical exponents from the Mayer cluster expansion coefficients: (i) construct the Matrix R, which is uniquely determined from the cluster integrals. (ii) verify that the matrix elements follow the asymptotic form (1) and determine the coefficients A,B, a and b. (iii) The critical exponents $\sigma$ and $\sigma'$, characterizing the non-analytic dependence of the density on z near the non-physical and physical singularities, respectively, are then simply given by

$$\sigma = \frac{1}{2}\sqrt{1 - 4\frac{2a+b}{A}} \qquad (2)$$

$$\sigma' = \frac{1}{2}\sqrt{1 - 4\frac{2a-b}{A}} \qquad (3)$$

(the sign of $\sigma$ and $\sigma'$ depends on the details of the matrix, see below). We have verified that $\sigma$, as predicted from (2), agrees with the expected universal values for a large number of models (see table 1). This is a strong support to the validity of our approach for the more physically relevant exponent $\sigma'$ as well.

We demonstrate the method by applying it to the well-studied hard spheres model. Using the 10 available virial coefficients [20], we show that the fluid branch analytically continues through the freezing density. It terminates with a singularity at density $\rho_t = 0.751(5)$, surprisingly close to the experimentally and computationally observed glass transition. The critical behavior of the density is found from (3) to be $\rho_{sing}(z) \sim (z_t-z)^{\sigma'}$, where $\sigma' = 0.0877(25)$.

In the following, we first provide evidence to the wide applicability of (1), then we sketch the derivation of equations (2-3), and finally we demonstrate the method by applying it to the hard-spheres model.

In contrast with the asymptotic behavior of the virial series, which for a variety of hard core lattice-gas models is dominated by a pair of non-physical complex singularities in the complex ρ plane [22], for systems with repulsive interactions the singularities of the cluster series are known to lie all on the real z axis [21]. Furthermore, the asymptotic form of the cluster coefficients of a repulsive system is well defined [24-26]:

$$nb_n \cong (-z_0)^{-(n-1)} n^{-\phi(d)} \{1 + cn^{-\theta(d)}\} \qquad (4)$$

The radius of convergence of the cluster series, $z_0$, is model-dependent, but the exponents $\phi(d)$ and $\theta(d)$ are universal, depending only on the spatial dimensionality d. It turns out [21] that a convenient representation of the equation of state can be obtained in terms of the real symmetric tri-diagonal matrix, R, defined by:

$$(R^n)_{11} = (-1)^n (n+1) b_{n+1}. \qquad (5)$$

The relation (5) uniquely determines the matrix R up to an irrelevant sign of the off-diagonal terms, see [22] for an explicit construction the matrix. The density is then given in terms of the R matrix by:

$$\rho(z) = \sum_{n=1}^{\infty} nb_n z^n = \sum_{n=0}^{\infty} (-1)^n z^{n+1} (R^n)_{11} = z(I + zR)^{-1}{}_{11}, \qquad (6)$$

where I is the identity matrix. The diagonal and near diagonal matrix elements rapidly converge to constant asymptotic values B and A, respectively. These values are related to the two branch points of the fluid thermodynamic functions, the non-physical universal singularity $z_0$ and the termination of the fluid branch $z_t$, by

$$z_0^{-1} = 2A + B \qquad z_t^{-1} = 2A - B \qquad (7)$$

Moreover, the leading correction for large n takes the form (1) for a large number of purely repulsive interaction models. Figure 1 presents the n dependence of $R_{nn}+2R_{n,n+1}$ for a few different repulsive interaction models, representing lattice and continuum models, hard-core as well as continuous, isotropic and anisotropic potentials, in varying spatial dimensionality. Clearly, the $1/n^2$ dependence sets in for n not too large. This asymptotic convergence was recently employed to yield accurate equations of states for the entire fluid branch, including the critical regime [27,28]. In Table 1 we list many more models in which the form (1) holds. The few exceptions are lattice gas models with nearest neighbors' exclusion (hard squares and hard hexagons), where the corrections to the asymptotic values A and B oscillates.

Since the asymptotic form (1) applies to a large set of models, we here study its implications to the critical behavior at the two singular points. Let R be a semi-infinite tridiagonal matrix, the non-zero elements of which are given by $R_{nn}=B+b/n^2$ and $R_{n,n\pm1}=A+a/n^2$; n=1,…∞. As our derivation below shows, analysis of the density in the vicinity of $z_0$ and $z_c$, requires knowledge of the eigenvalues and eigenfunctions of R near the spectrum edge. The eigenvalue equations for R read:

$$(A+a/n^2)(\psi_{n-1}+\psi_{n+1})+(B+b/n^2-\lambda(k))\psi_n=0 \qquad (8)$$

with the boundary condition $\psi_0=0$. The unperturbed matrix (a=b=0) has a continuous spectrum $\lambda(k)=B+2A\cos(k)$, where $0\leq k<\pi$, and the corresponding eigenfunctions are $\psi_n=\sqrt{\frac{2}{\pi}}\sin(kn)$. If the perturbation is weak enough, the spectrum does not change under the perturbation, and only the eigenfunctions are modified. We are mostly interested in the extremal eigenvalues, corresponding to k<<1 or π-k<<1, where the eigenfunction vary slowly with the index n, and the continuum limit is appropriate (up to a sign change for k~π, see below). We present the analysis for k~0.

Denote $\psi_n=f(kn)$, where f is to be determined and k<<1. Expanding $\psi_{n\pm1}$ And $\lambda(k)$ to second order in k, and denoting x=kn, one obtains the following equation for f(x)

$$f''(x)+\frac{((b+2a))/x^2+A}{A}f(x)=0 \qquad f(0)=0 \qquad (9)$$

whose solution is $f(x) = x^{1/2} J_\sigma(x)$ where $J_\sigma(x)$ is the Bessel function of order $\sigma$, and $\sigma$ is given by equation (2). For a=b=0 one recovers the sin(kx) solution.

According to Eq. (6), the density can be written as $\rho(z) = \int_0^\pi \frac{u(k)}{z^{-1} + \lambda(k)} dk$, where u(k) is just $|\psi_1(k)|^2 = f^2(k)$. For z near $-z_0 = -(B+2A)^{-1}$, the singular contribution to the integral comes from k~0, where $u(k) \sim k^{2\sigma+1}$. Asymptotic expansion of the integral thus yields

$$\rho(-z_0) - \rho(z) \sim C(z+z_0)^\sigma. \tag{10}$$

The above analysis applies to matrices described by (1). In all physical models we considered, this behavior sets in approximately for n>2, but the first matrix elements, especially $R_{11}$ and $R_{12}$, deviate significantly from the asymptotic form. A convenient way to correct for this is through the continued fraction representation for the density

$$\rho(z) = \frac{1}{R_{11} + 1/z - R_{12}^2 \tilde{\rho}(z)} \tag{11}$$

where $\tilde{\rho}(z)$ is the density one would have calculated from the matrix R after removing its first row and column. One may find $\tilde{\rho}(z)$ for the truncated matrix which does satisfy the form (1), and obtain the density by successive application of (11). The singular part of the density $\rho(z)$ is just that of $\tilde{\rho}(z)$, unless the denominator in (11) vanishes at the critical activity. In the latter case, the density diverges at $z=-z_0$: $\rho(z) \sim C(z+z_0)^{-\sigma}$. In all 1D and 2D models we studied, such cancellation happens in the vicinity of $-z_0$. For d≥3, there is no such cancellation, and the critical exponent is $\sigma>0$. In both cases, this exponent, determined by the matrix elements behavior, is just the universal exponent $\phi(d)-1$ [26]. Thus, the critical exponent $\phi(d)$ is determined by the convergence of the matrix elements towards their asymptotic values, and can be easily extracted from the n-dependence of the first few matrix elements. Using the known values of $\phi(d)$ one obtains the universal relations

$$\frac{2a+b}{A} = \begin{cases} 0 & d=1 \\ 2/9 & d=2 \\ 0.2424(4) & d=3 \\ 0.180(1) & d=4 \\ 0.088(4) & d=5 \\ 0 & d \geq 6 \end{cases} \qquad (12)$$

which are indeed satisfied for all 14 models studied (see Table 1).

As the asymptotics at k~0 successfully reproduces the correct universal behavior, one is encouraged to apply the same approach to the more interesting non-universal singularity at $z_t=(2A-B)^{-1}$, characterizing the physical termination point of the fluid phase. Here one uses $\psi_n=(-1)^n f(kn)$, and the singular contribution to the integral comes from k~π. The same differential equation follows (except for a sign change of A), and the singular behavior near the physical singularity of the fluid branch is described by ($z_t<\infty$)

$$\rho(z_t)-\rho(z) \sim C'(z_t-z)^{\sigma'} \qquad (13)$$

where σ' is given by (3). Accordingly, the critical behavior of the equation of state is

$$p(z_t)-p \sim (\rho(z_t)-\rho)^{1/\sigma'} \qquad (14)$$

If $z_t=\infty$ (e.g., for the hard-lines model) one obtains $\rho(z=\infty)-\rho(z) \sim C'z^{-\sigma'}$ and a logarithmic divergence of the pressure at the transition. Again, cancellation of the denominator in (11) changes the sign of the exponent, as indeed happens for the Gaussian model, where the density diverges at $z_t=\infty$.

In summary, we presented here a simple method to predict the critical behavior at the termination of the fluid phase from the behavior of the Mayer cluster integrals. These converge quite rapidly for many models, allowing for an accurate prediction for the critical exponent.

As an example, we apply the above method to the 3D hard spheres model. The virial coefficients $B_n$ for hard spheres were recently calculated up to n=10 [20]. The resulting R matrix elements are presented in table 2 (evaluation of $R_{5,6}$ uses the extrapolated value of $B_{11}$ [20]). Figure 2 presents the convergence of the matrix elements towards the asymptotic values A=2.74553(3) and B=5.49095(6). One clearly sees that for the hard spheres model the observed correction for the diagonal terms $R_{nn}$ scales like $1/n^3$, such that the $1/n^2$ term, in exists, is very weak. We thus conclude that in this case b≈0. It thus follows from eqs. (2,3) we find that σ'=σ, namely the critical exponent at the fluid phase termination point is the same as the universal exponent σ'=σ=φ(3)-1=0.0877(25). Consistently, the off-diagonal terms are well fitted by $R_{n,n+1}=A(1+0.121/n^2)$ (see figure 2), in agreement with equation (12) and b≈0. Note that using the above approach one can predict the critical exponent even if the transition activity itself can not be accurately extracted.

The fluid phase terminates at $z_t=(2A-B)^{-1}$. Since 2A and B are very close, we cannot give an accurate estimate of the transition activity. However, the critical density is insensitive to the exact value of $z_t$, and we find that $\rho_t$=0.751(5). A similar insensitivity of the fluid branch to its termination point $z_t$ has been observed for the two dimensional N3 lattice model [27] and the hard disks model [28]. Note that the virial series analysis has not provided any evidence for singularity at $\rho_t$, presumably due to the screening effect of the leading complex singularities, as explained in [20]. Analysis of the virial series in other lattice models does not detect the physical singularity as well [20]. The termination density $\rho_t$ is well above the freezing density $\rho_f$=0.663(2) and thus we conclude that the fluid branch is analytic at the freezing density, and can be continued to describe a metastable super-cooled fluid. This super-cooled fluid phase is thus expected to terminate at the critical density $\rho_t$, which is surprisingly close to the density at which the glass transition was experimentally observed, and is much lower than the random closest packing density $\rho_{rcp}$~0.87 [29]. Close to the transition, the pressure is predicted to either increase rapidly followed by a flat approach to $p_t$ (Eq. (14) with 1/σ' = 11.4(4)) for $z_t<\infty$, or to diverge logarithmically ~-ln($\rho_c-\rho$) for $z_t=\infty$. The latter scenario seems more compatible with present MC and molecular dynamics data. We thus conjecture that the termination point of the metastable super-cooled fluid phase obtained here could be the singularity characterizing a glass transition of the hard spheres.

In conclusion, we show that the matrix representation of the Mayer cluster integrals posses a well-defined asymptotic behavior for a large class of repulsive interaction models. Using this form, we find a general exact expression for the critical exponents of such systems in both the unphysical and physical singularities on the real activity axis. For all 14 models tested, the results for the unphysical singularity are consistent with the known universal value of the critical exponent, thus providing a strong support to the validity of our approach for the physical singularity as well. The method is applied to the hard spheres model, where we conclude from the available 10 virial coefficients that the fluid phase terminates at $\rho_t=0.751(5)$, very close to the observed glass transition, and the critical exponent is predicted to be $\sigma'=0.0877(25)$.

## ACKNOWLEDGMENTS

We thank M. Schwartz for critical reading of the manuscript. E.E. acknowledges support from the Alon fellowship at Tel-Aviv University.

**Figure Legends:**

**Figure 1**: The asymptotic convergence of elements of the tri-diagonal matrix R, $R_{nn}+2R_{n,n+1}$ as a function of n for different models: hard lines in continuous one-dimension, square lattice 2D model with next-nearest neighbor exclusion, dimers on the 3D cubic lattice, and the Gaussian model in 4D. The symbols are the exact values and the lines are best fits to the $1/n^2$ behavior, except for the 1D case which is fitted well by a $1/n^3$ curve since the coefficient of the $1/n^2$ term vanishes, as expected from universality (see equation (12)).

**Figure 2**: Convergence of the diagonal $R_{nn}$ (circles) and twice the off-diagonal $2R_{n,n+1}$ (triangles) elements of the tri-diagonal matrix R for the hard-spheres model towards their respective asymptotic values B and A. The corrections for the off-diagonal and diagonal terms are well-fitted by $.332/n^2$ and $0.083/n^3$ respectively (error bars are not visible, as they are smaller than symbol sizes).

# Figures:

# Figure 1:

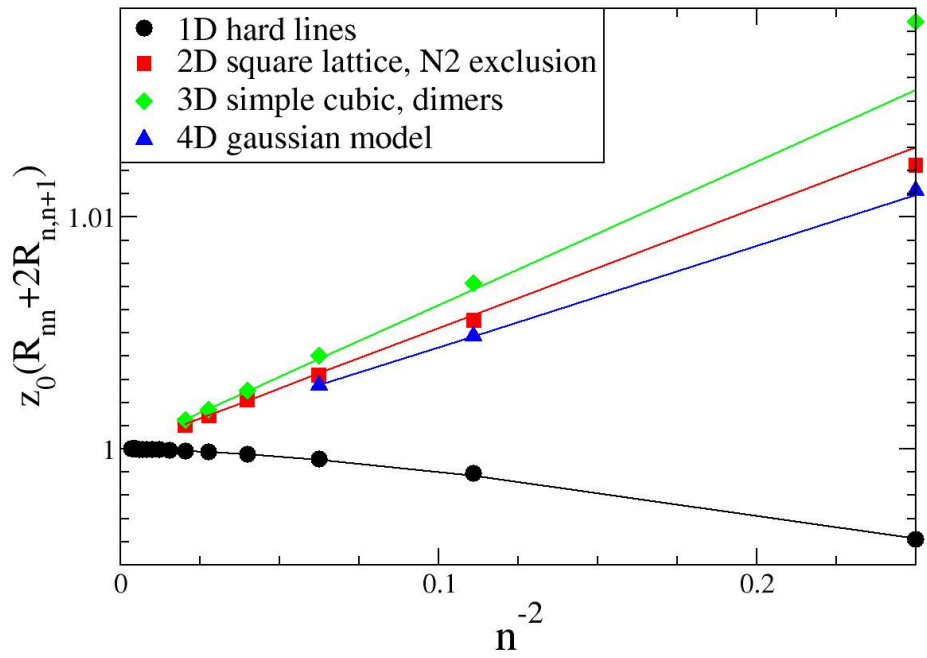

**Figure 2**:

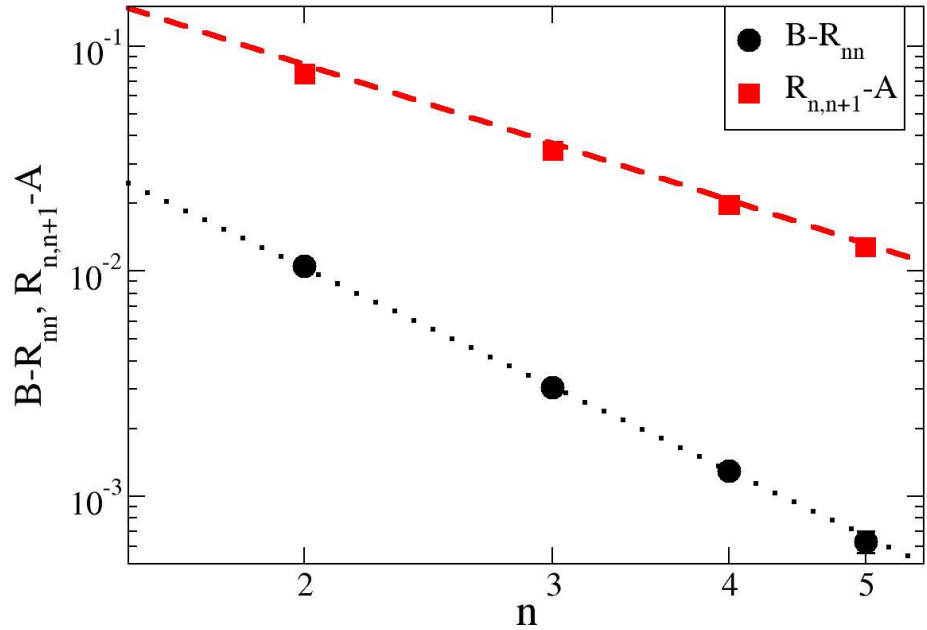

**Table 1:** Parameters obtained from fitting the R matrix elements to the form (1), for several repulsive interaction models. In all these examples, the matrix elements follow the $1/n^2$ convergence, and the coefficients obtained are in agreement with the values predicted based on the universality of the negative-activity singularity [5].

| Model | d | Number of available coefficients | A | B | (2a+b)/A (fit of matrix elements) | (2a+b)/A (expected from universality [22]) |
|---|---|---|---|---|---|---|
| Hard lines | 1 | Exact | e/4 | e/2 | 0 | 0 |
| Hard disks[20] | 2 | 10 | 1.4839(2) | 2.9675(3) | 0.212(10) | 2/9 |
| Gaussian model [2&3] | 2 | 10 | .9637(2) | 1.9274(2) | 0.242(5) | |
| Square Lattice, 2$^{nd}$ neighbor exclusion [30] | 2 | 16 | 3.6312(2) | 7.2598(2) | 0.222(3) | |
| Square Lattice, 3$^{rd}$-neighbor exclusion [26] | 2 | 13 | 5.374(1) | 10.7426(5) | 0.227(5) | |
| Triangular Lattice, 2$^{nd}$-neighbor exclusion [31] | 2 | 10 | 5.3974(5) | 10.5990(5) | 0.225(2) | |
| Dimers, triangular Lattice [32] | 2 | 15 | 4.4584(4) | 8.9162(6) | 0.230(8) | |
| Hard spheres[20] | 3 | 10 | 2.7455(1) | 5.4909(1) | 0.2423(2) | 0.2424(4) |
| Gaussian model [2&3] | 3 | 10 | 1.114(1) | 2.228(1) | 0.253(5) | |
| Dimers, Simple Cubic Lattice [32] | 3 | 15 | 4.809(1) | 9.618(2) | 0.265(20) | |
| Dimers, bcc Lattice [32] | 3 | 13 | 6.7017(15) | 13.3970(10) | 0.247(6) | |
| Gaussian model [2&3] | 4 | 10 | 1.2155(5) | 2.4310(5) | 0.1802(5) | 0.180(1) |
| Gaussian model [2&3] | 5 | 10 | 1.2788(5) | 2.5576(5) | 0.082(1) | 0.088(4) |
| Gaussian model [2&3] | 6 | 10 | 1.3154(1) | 2.6108(1) | 0.001(1) | 0 |

**Table 2:** First 10 non-vanishing matrix elements of the tridiagonal R matrix for the hard-spheres model, based on the first 10 virial coefficients[20].

| n | 1 | 2 | 3 | 4 | 5 |
|---|---|---|---|---|---|
| $R_{nn}$ | 5.9238439 | 5.4804133 | 5.4878894(70) | 5.4896498(250) | 5.4903217(700) |
| $R_{n,n+1}$ | 3.0530793 | 2.8206978(5) | 2.7797594(70) | 2.7651616(200) | 2.7582884(600) |